\documentclass{elsart}
\usepackage{graphicx}

\begin{document}

\begin{frontmatter}

\title{Measurements of Atmospheric Antiprotons}

\author[kobe]{K.~Yamato},
\author[kobe] {K.~Abe\thanksref{icrr}},
\author[isas]  {H.~Fuke},
\author[kek]  {S.~Haino},
\author[kek]  {Y.~Makida},
\author[tokyo]{S.~Matsuda},
\author[tokyo]{H.~Matsumoto},
\author[gsfc] {J.W.~Mitchell},
\author[gsfc] {A.A.~Moiseev},
\author[tokyo]{J.~Nishimura},
\author[kobe] {M.~Nozaki},
\author[tokyo]{S.~Orito\thanksref{dead}},
\author[gsfc] {J.F.~Ormes},
\author[tokyo]{\underline{T.~Sanuki}\corauthref{cor1}},
\corauth[cor1]{Corresponding author.}
\ead{sanuki@icepp.s.u-tokyo.ac.jp}
\author[gsfc] {M.~Sasaki},
\author[umd]  {E.S.~Seo},
\author[kobe] {Y.~Shikaze\thanksref{jaeri}},
\author[gsfc] {R.E.~Streitmatter},
\author[kek]  {J.~Suzuki},
\author[kek]  {K.~Tanaka},
\author[isas] {T.~Yamagami}, 
\author[kek]  {A.~Yamamoto},
\author[kek]  {T.~Yoshida},
\author[kek]  {K.~Yoshimura}

\address[kobe] {Kobe University, Kobe, Hyogo 657-8501, Japan}
\address[isas] {The Institute of Space and Astronautical Science (ISAS)
                of Japan Aerospace Exploration Agency (JAXA),
                Sagamihara, Kanagawa, 229-8510, Japan}
\address[kek]  {High Energy Accelerator Research Organization (KEK),
                Tsukuba, Ibaraki 305-0801, Japan}
\address[tokyo]{The University of Tokyo, Bunkyo, Tokyo 113-0033, Japan}
\address[gsfc] {National Aeronautics and Space Administration (NASA),
                Goddard Space Flight Center (GSFC),
                Greenbelt, MD 20771, USA.}
\address[umd]  {University of Maryland, College Park, MD 20742, USA}
\thanks[icrr]{Present address: ICRR, The University of Tokyo, 
              Kashiwa, Chiba 227-8582, Japan}
\thanks[dead]{deceased.}
\thanks[jaeri]{Present address:  JAEA, Tokai, Ibaraki 391-1195, Japan}

\begin{abstract}
We measured atmospheric antiproton spectra 
in the energy range 0.2 to 3.4 GeV, 
at sea level
and at balloon altitude in the atmospheric depth range 4.5 to 26 g/cm$^2$.
The observed energy spectra,
including our previous measurements at mountain altitude,
were compared with estimated spectra
calculated on various assumptions regarding the energy distribution of antiprotons 
that interacted with air nuclei.
\end{abstract}

\end{frontmatter}

% -------------------------------------------------------------------
% I. Introduction
% -------------------------------------------------------------------
\section{Introduction}
\label{sec:introduction}
Antiprotons are produced in the atmosphere via interaction 
between cosmic rays and atmospheric nuclei.
These atmospheric antiprotons are carrying important information 
about the physical processes of their production and propagation 
in the atmosphere. 
The production and propagation mechanisms should be similar 
to those of galactic antiprotons. 
Therefore the study of atmospheric antiprotons is important 
for understanding the energy spectrum of galactic antiprotons. 
%for understanding galactic antiprotons. 
In addition atmospheric antiprotons are a major background 
for the galactic antiprotons measured at balloon altitude. 
This background has to be estimated by model calculations. 
However model calculations of this background%
~\cite{bib:BowenMoats,bib:Pfeifer,bib:Mitsui,bib:Stephens1997,bib:HuangDth,bib:Huang}
have not yet been verified by direct observation. 

We report here the atmospheric antiproton spectra 
measured with the BESS detector 
in a kinetic energy region of 0.2 -- 3.4 GeV 
at sea level at the atmospheric depth 994 g/cm$^2$ 
in 1997 (BESS-1997), 
and at balloon altitude 
over the atmospheric depth range 4.5 -- 26 g/cm$^2$ in 2001 (BESS-2001). 
The antiprotons observed at balloon altitude are assured not to be primary, 
but are produced inside the atmosphere, 
because the vertical geomagnetic cut-off energy was 3.4 GeV for protons/antiprotons 
throughout the balloon flight.
The measured spectra,
including our previous data 
observed at mountain altitude at the atmospheric depth 742 g/cm$^2$  
in 1999 \cite{bib:Sanuki_norikura},
will be compared with two model calculations.
We will discuss how these calculations provide fits to the observed data.

% -------------------------------------------------------------------
% II. BESS Spectrometer
% -------------------------------------------------------------------
\section{The BESS spectrometer}
\label{sec:detector}

The detector 
for the Balloon-borne Experiment with a Superconducting Spectrometer (BESS)
was designed~\cite{bib:BESS1,bib:BESS2} and developed~
\cite{bib:BESS3,bib:BESS4,bib:BESS5,bib:BESS6} as a high-resolution
magnetic-rigidity spectrometer with a large acceptance 
to perform highly sensitive searches for rare cosmic-ray components,
as well as precise measurements of
various cosmic-ray species~\cite{bib:Asaoka-pbar,bib:Sanuki-p,bib:Abe}.
A uniform magnetic field of 1 Tesla is produced by a thin superconducting
solenoid~\cite{bib:BESS_MAG}. The magnetic-rigidity ($R \equiv Pc/Ze$)
of an incoming charged particle is measured by a tracking system
which consists of a jet-type drift chamber and two inner drift chambers
inside the magnetic field. The deflection ($R^{-1}$) is calculated for
each event by applying a circular fit using up to 28 hit points,
each with a spatial resolution of 200 $\mu$m.
Upper and lower scintillator hodoscopes~\cite{bib:BESS6} provide
time-of-flight and two independent d$E$/d$x$ measurements. 
Time resolution of each counter is 55 ps, 
resulting in a 1/$\beta$ resolution of 1.4\%, where
$\beta$ is defined as the particle velocity normalized by the speed of light. 
The first-level trigger is provided 
by a coincidence between the top and the bottom scintillators 
with the threshold set at 1/3 of the pulse height for minimum ionizing particles.
In the BESS-1997 ground experiment, all the first-level triggered events were recorded
because the trigger rate was 30 Hz,
which is low enough to record all the events.
In the BESS-2001 balloon-flight experiment, 
the first-level trigger rate was too high
to record all the events,
and thus 
a second-level trigger was issued 
when a particle's rigidity was calculated to be negative 
by the onboard computer \cite{bib:BESS4,bib:Maeno-pbar}
to record negatively charged particles preferentially. 
In addition to this biased trigger mode,
one out of every four first-level triggered events were recorded 
to provide an unbiased data sample in the balloon flight. 
The BESS instrument also incorporates a threshold-type Cherenkov counter \cite{bib:BESS5},
whose radiator was a silica-aerogel 
with a reflective index of 1.03 in BESS-1997 and 1.02 in BESS-2001. 
Antiprotons are distinguished from $e^-$ and $\mu^-$ background 
by imposing that there be no light output from the Cherenkov counter. 
The rejection factor in this analysis was about 50,000 and 7,000 
in BESS-1997 and BESS-2001, respectively. 

% -------------------------------------------------------------------
% III. Observations1
% -------------------------------------------------------------------
\section{Observations}
\label{sec:observations}
\begin{figure}[ht]
  \begin{center}
    \includegraphics[width=10.0cm]{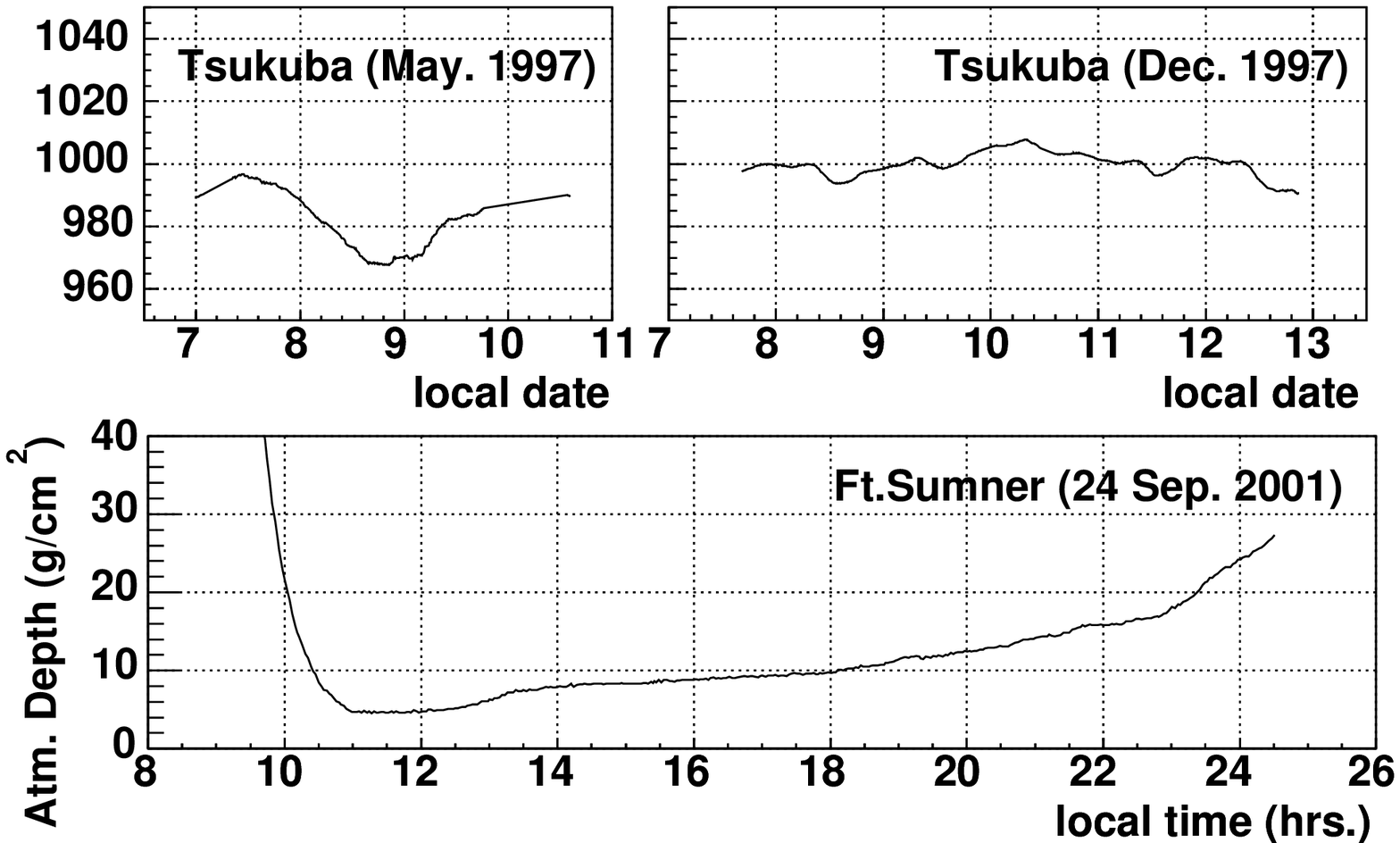}
    \caption{Atmospheric depth during the BESS-1997 ground observations
     and the BESS-2001 balloon flight.}
    \label{fig:press}
  \end{center}
\end{figure}

The BESS-1997 observations at sea level were carried out 
at the High Energy Accelerator Research Organization (KEK), Tsukuba, Japan 
(36$^\circ$ 12$'$ N, 140$^\circ$ 6$'$ E), where the geomagnetic
cut-off rigidity was 11.2 GV,
during the two periods of 6th -- 11th May, and 7th -- 13th December in 1997. 
The BESS-2001 balloon flight was carried out at Ft. Sumner, New Mexico,
USA (34$^\circ$ 49$'$ N, 104$^\circ$ 22$'$ W) on 24th September 2001.
Throughout the flight, the vertical geomagnetic cut-off rigidity was
about 4.2 GV, which corresponds to a kinetic energy of 3.4 GeV for protons/antiprotons.
After the balloon reached a normal floating altitude of 37 km,
where the atmospheric depth is 4.5 g/cm$^2$,
it began to lose altitude
and continued descending for 13 hours 
before the termination of the flight at an atmospheric depth of around 30 g/cm$^2$. 
The atmospheric depths throughout the observations are shown in Fig. \ref{fig:press}.
The mean atmospheric depth during sea level and balloon altitude observations was 
994 g/cm$^2$ and 10.7 g/cm$^2$, respectively.

% -------------------------------------------------------------------
% IV. Data analysis
% -------------------------------------------------------------------
\section{Data analysis}
\label{sec:analysis}

\begin{figure}[ht]
  \begin{center}
    \includegraphics[width=12.0cm]{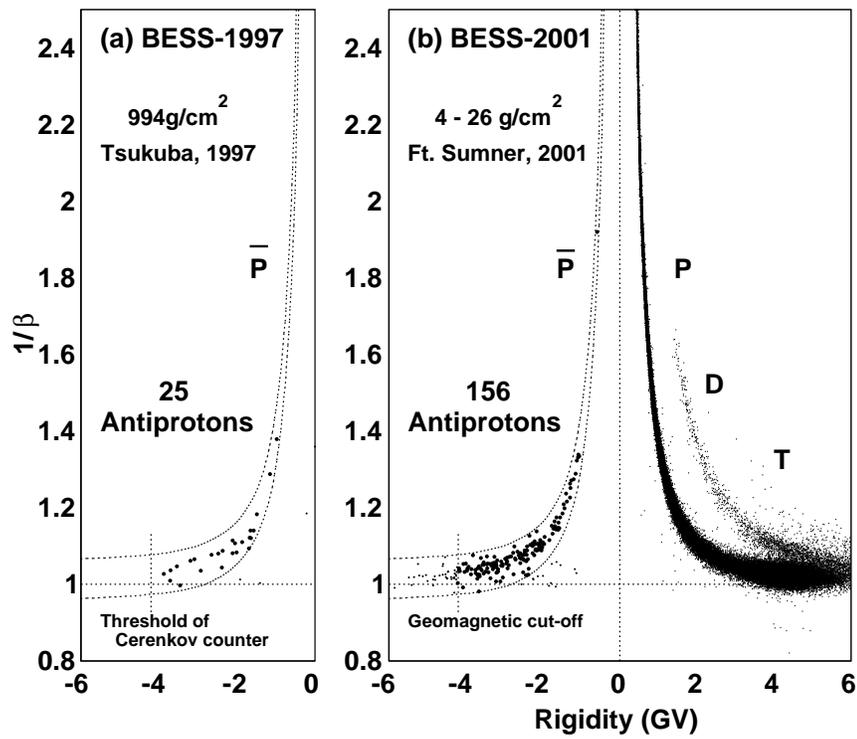}
    \caption{The identification plots of antiproton events for 
    (a) the BESS-1997 observations at sea level, and 
    (b) the BESS-2001 balloon flight.
    The dotted curves define the antiproton mass bands.}
    \label{fig:idplot}
  \end{center}
\end{figure}

Analysis was performed in the same way as in the previous BESS
experiments~\cite{bib:Asaoka-pbar,bib:Orito-pbar}.
We first selected events 
without interactions inside the BESS detector and
with good-quality measurements of the rigidity and velocity.
Particle identification was performed by requiring that 
d$E$/d$x$ be consistent with a singly charged particle,
no signal was observed from the silica-aerogel Cherenkov counter, and 
particle mass was calculated to be consistent with the antiproton mass. 
In Fig. \ref{fig:idplot}, events between the two curves were
identified as antiprotons.
The number of detected antiproton candidates in the energy range 0.2  - 3.4 GeV were
25 at sea-level in BESS-1997 
and 156 at balloon altitude in BESS-2001.

In order to obtain the absolute flux of antiprotons at the top of the BESS
instrument, we estimated the event selection efficiency, interaction
loss probability and energy loss inside the instrument, and background
contamination. The selection efficiency was obtained by using the recorded data, 
which consist mainly of protons since d$E$/d$x$ and 1/$\beta$
are the same as those of antiprotons. The interaction probability and energy
loss inside the instrument, as well as its geometrical acceptance, were
calculated by Monte Carlo (MC) simulation. The MC code was tuned and
verified by comparing the simulation with an accelerator beam test of
the BESS detector~\cite{bib:BeamTest}.
The systematic uncertainty of the event selection efficiency and antiproton
interaction losses were evaluated to be 5\% from this beam test.
The background contamination due to inefficiency of the silica-aerogel
Cherenkov counter was estimated by dividing $N^{-}$ by the rejection
factor of the Cherenkov counter, where $N^{-}$ is the number of negatively charged
particles after applying all selection cuts except for the Cherenkov veto to
the entire set of events. The rejection factor was estimated using the
recorded proton events during the balloon flight. In the energy range
between 1.9 and 3.4 GeV, the background contamination was found to be 
20\% and 5\% 
in BESS-1997 and BESS-2001, respectively.
Between 1.0 and 1.9 GeV, 
the contamination was less than 1\% for both observations.
Below 1.0 GeV, it was negligibly small.

\begin{figure}[ht]
  \begin{center}
    \includegraphics[width=12.0cm]{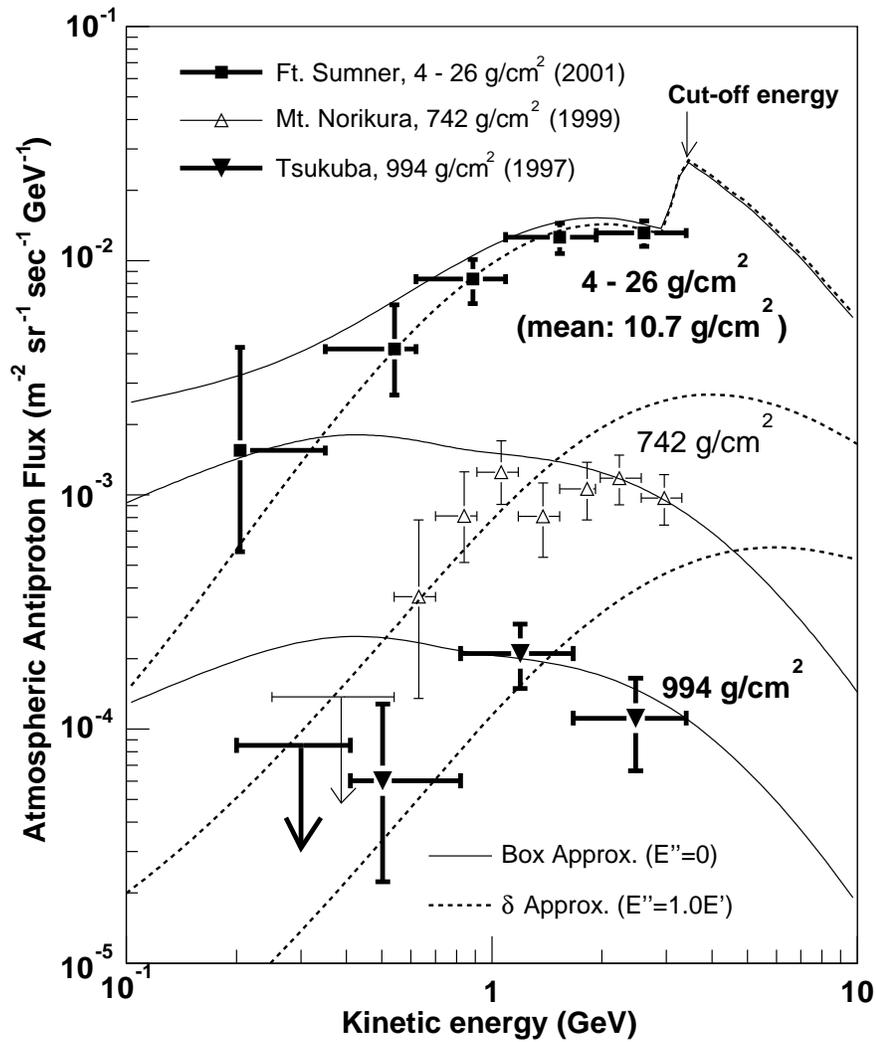}
    \caption{The observed antiproton flux 
    %at atmospheric depth of 4.5 - 26 g/cm$^2$ at Ft. Sumner, 2001
    at atmospheric depth of 4.5 - 26 g/cm$^2$ at Ft. Sumner in 2001
    and
    %at 994 g/cm$^2$ at Tsukuba, 1997.
    at 994 g/cm$^2$ at Tsukuba in 1997.
    The flux at 742 g/cm$^2$ at Mt. Norikura~\cite{bib:Sanuki_norikura} is also shown.
    The results are compared with calculations assuming box-approximation (solid lines)
    and $\delta$-approximation (dashed lines).
    }
    \label{fig:flux}
  \end{center}
\end{figure}

Fig. \ref{fig:flux} shows the resultant antiproton flux observed 
at sea level in BESS-1997 and 
at balloon altitude in BESS-2001.
These results are summarized in Tables \ref{tab:flux_gnd} and \ref{tab:flux_e34}.
The detectable energy range was limited 
by the threshold of the silica-aerogel Cherenkov counter's ability 
to reject background in BESS-1997.
As for the BESS-2001 experiment,
the upper limit of the energy range was set 
not to exceed the geomagnetic cut-off energy. 
Results of the model calculations discussed below, 
as well as our previously published data obtained at mountain altitude  \cite{bib:Sanuki_norikura},
are also shown in Fig. \ref{fig:flux}.

% -------------------------------------------------------------------
% V. Model calculations
% -------------------------------------------------------------------
\section{Model calculations}
\label{sec:calc}
Several calculations of the atmospheric antiproton flux have been published%
~\cite{bib:BowenMoats,bib:Pfeifer,bib:Mitsui,bib:Stephens1997,bib:HuangDth,bib:Huang}.
We made a phenomenological calculation of atmospheric antiproton flux
under the observation conditions, 
following those model calculations. 
Referring to the transport equation given by Stephens~\cite{bib:Stephens1997}, 
we solved the following equation:
\begin{eqnarray}
  \frac{\partial J_{\bar{p}}(E,x) }{\partial x}
    = \sum_A Q_A(E_A, x, E)
  + \frac{\partial}{\partial E}
    \left[ J_{\bar{p}}(E,x)
    \left< \frac{\mbox{d}E}{\mbox{d}x} \right> \right]
  - \frac {J_{\bar{p}}(E,x)} {\Lambda(E)} \nonumber \\
  + \int_E^{\infty} \Phi(E, E')
    \left[(1-\alpha) \frac{J_{\bar{p}}(E',x)}{\lambda^{in}(E')}
    + \alpha \frac{J_{\bar{n}}(E',x)}{\lambda^{in}(E')}
    \right] \mbox{d}E' 
  \label{eq:stephX}
\end{eqnarray}
Here, $J_{\bar{p}}(E,x)$ is the differential antiproton flux at
the atmospheric depth $x$ g/cm$^2$. 
The first term on the right-hand side, $Q_A(E_A, x, E)$, represents the production rate of antiprotons
with energy $E$ by the incident particle $A$ with energy $E_A$.
Since the flux of incident particles depends on the atmospheric depth $x$,
$Q_A( E_A, x, E)$ is a function of $x$.
The second term represents the flux change due to the ionization energy loss.
The average energy loss per g/cm$^2$ is indicated by
$\left<\mathrm{d}E/\mathrm{d}x\right>$.
The third term represents loss of antiprotons due to interactions. 
$\Lambda(E)$ is the total inelastic interaction mean free path (mfp). 
It is described as $\Lambda(E) = [1/\lambda^{ann}(E) + 1/\lambda^{in}(E)]^{-1}$,
where $\lambda^{ann}(E)$ and $\lambda^{in}(E)$ are mfp's for
annihilating  and non-annihilating processes of antiprotons 
passing through the atmosphere, respectively. 
The fourth integral term represents the tertiary antiproton production rate
by non-annihilative inelastic interaction. 
$\Phi(E, E')$ signifies the probability that
an antiproton with initial energy of $E'$ possesses energy $E$ after a collision.
The charge exchange probability between antiprotons and antineutrons is described as
$\alpha$, which is taken to be 1/3.
Therefore $\Phi(E,E') (1-\alpha) J(E')_{\bar{p}} / \lambda^{in}(E')\, dE'$ represents
the production rate of tertiary antiprotons with energy of $E$ 
by the parent antiproton with an energy between $E'$ and $E'+dE'$.
In addition to the tertiary production from antiprotons, a contribution
from antineutrons ($\bar{n}$) is also included in this integral term.

The primary cosmic-ray flux used in this calculation was based on the results from
the BESS-98~\cite{bib:Sanuki-p} and AMS~\cite{bib:AMS} experiments,
the results of which are in good agreement. %with each other. 
The proton, neutron, and helium fluxes at various atmospheric depths were 
obtained by using the transport equation, 
and a contribution from heavier nuclei such as CNO was included, 
%by multiplying helium flux by a factor of 1.25, 
following Papini et al.~\cite{bib:Papini}.
The antiproton production spectrum by the cosmic rays was 
taken from Stephens's formulation~\cite{bib:Stephens1997}.

\begin{figure}[ht]
  \begin{center}
    \includegraphics[width=12.0cm]{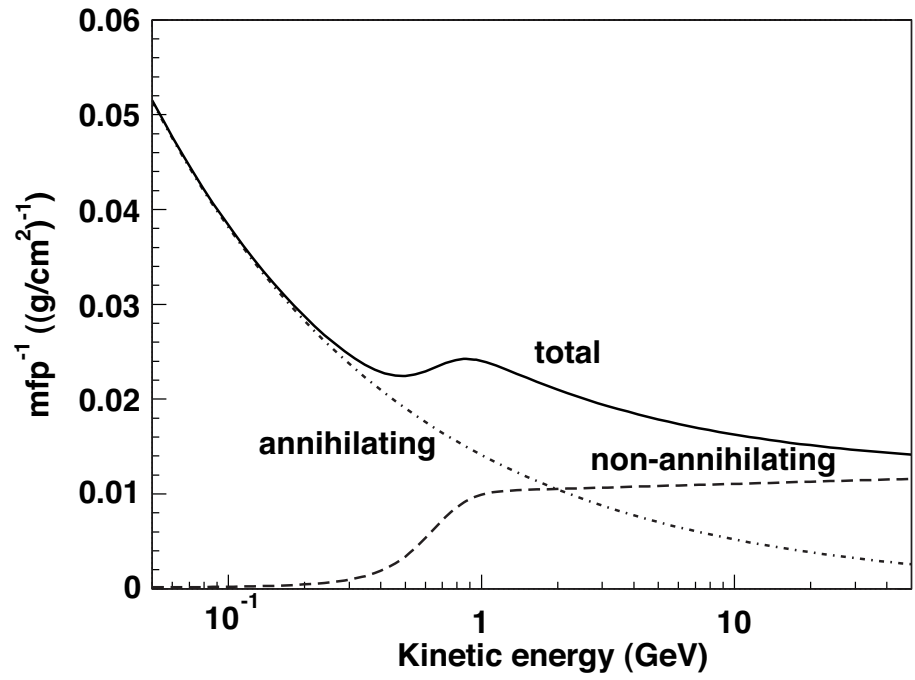}
    \caption{Mean free paths assumed in the calculations.}
    \label{fig:lambda}
  \end{center}
\end{figure}

The mfp's, $\Lambda(E)$,  $\lambda^{in}(E)$ and $\lambda^{ann}(E)$, are the same
as those adopted by Stephens~\cite{bib:Stephens1997}.
These are shown in Fig.~\ref{fig:lambda}.
The details of the parameters are explained by
Stephens~\cite{bib:Stephens1997} and by Tan and Ng~\cite{bib:TanNg}.

Since there is no direct experimental data on antiproton energy distribution 
after collision with the air target,
we examined several forms of $\Phi(E, E')$ in this study. 
One was very similar to those adopted by 
Bowen and Moats~\cite{bib:BowenMoats}, Stephens~\cite{bib:Stephens1997} 
and Tan and Ng~\cite{bib:TanNg}.
In this model, 
$\Phi(E,E')$ was assumed to be of the form of $1/E' (0 \leq E \leq E')$, 
which means the probability that
an antiproton with initial energy $E'$ possesses energy $E$ 
after a collision is uniform from $E = 0$ to $E = E'$. 
The average energy after a collision is half of the initial energy. 
We call this model a ``box-approximation'', 
because the energy distribution of produced tertiary antiprotons has a box-shaped spectrum.
Another model was proposed by Huang on a different assumption \cite{bib:HuangDth}. 
In his evaluation of total inelastic interaction, 
only annihilation channels were taken into account in the inelastic interactions.
Non-annihilating inelastic processes were
not included as a process of antiproton energy loss.
This assumption corresponds to taking $\Phi(E,E') = \delta(E'-E)$. 
This is an extreme case in which a tertiary antiproton does not loose its energy 
in a collision with an atmospheric nucleus.
We call this model a ``$\delta$-approximation''.
Other intermediate models between the above two models could be considered,
such as $\Phi(E,E') = 1/(E'-E'') (E'' \leq E \leq E')$, where $E''$ is taken to be 
various between 0 and $E'$. 
The box- and $\delta$-approximations correspond to taking $E''$ as 0 and $E'$, respectively.

In order to compare the observed flux with the calculated results, the following
two effects were taken into account:
the dependence of the geometrical acceptance and antiproton flux on the zenith angle, 
and the effective observation time at each atmospheric depth. 
The weighted averaged flux taking into account these points, $\left< J_{\bar{p}}(E) \right>$, was
defined as
\begin{equation}
  \left< J_{\bar{p}}(E) \right>  \equiv 
    \frac{\int_{0}^{+1} \mathrm{d}(\cos\theta)
    \int_{x_1}^{x_2} \mathrm{d}x
    J_{\bar{p}}(E, x^{\ast}(x,\theta) )
    \Delta T(x_1, x_2)
    \Delta S\Omega(E, \cos\theta)}
    {\int_{0}^{+1} \mathrm{d}(\cos\theta)
    \int_{x_1}^{x_2} \mathrm{d}x
    \Delta T(x_1, x_2)
    \Delta S\Omega(E, \cos\theta)},
  \label{eq:weight_ave}
\end{equation}
where $J_{\bar{p}}$ is a solution of Eq. (\ref{eq:stephX}) 
and $x^{\ast}$ denotes the effective atmospheric depth,
which depends on zenith angle of incoming parent particles.
$\Delta T(x_1,x_2)$ is the live observation time between atmospheric depth $x_1$ and $x_2$.  
$\Delta S\Omega(E,\cos\theta)$ is the geometrical acceptance 
corresponding to the energy and range of zenith angle, obtained by the MC simulation. 
In the calculation for balloon altitude, the zenith angle dependence
of antiprotons was assumed to be $x^{\ast}(x,\theta) = x / \cos \theta$,
because the number of antiprotons produced should be proportional
to the path length of the primary parent particle.
On the other hand, 
since the tertiary antiprotons dominate in those observed at large atmospheric depth, 
they might have collided with nuclei several times and changed their directions. 
In this case, their flux is very sensitive to the mfp inside the atmosphere,
and their real path length cannot be inferred from the observed zenith angle. 
We simply assumed $x^{\ast}(x,\theta) = x$ in this study.
This assumption does not significantly change the antiproton spectral shape,
but may change its absolute flux at a large atmospheric depth.

% -------------------------------------------------------------------
% VI. Results and discussions
% -------------------------------------------------------------------
\section{Results and discussions}
\label{sec:results}

The resultant energy spectra of antiprotons 
observed at sea level (BESS-1997) and at balloon altitude (BESS-2001)
are shown in Fig. \ref{fig:flux}.
These results are compared with 
our previous data observed at Mt. Norikura, Japan \cite{bib:Sanuki_norikura},
where the mean atmospheric depth was 742 g/cm$^2$,
together with the results of the corresponding model calculations.
Since the primary galactic antiproton flux was taken into account in the calculation, 
a sharp edge is seen at the cut-off energy of around 3.5 GeV 
in the calculated flux at 4.5 -- 26 g/cm$^2$ at Ft. Sumner. 
Here, we adopted the galactic antiproton spectrum calculated by
Mitsui~\cite{bib:Mitsui} with the solar activity characterized by $\Phi=1000$~MV.
The antiprotons observed at balloon altitudes are assured to be purely atmospheric 
because the highest energy of the measurement
is below the geomagnetic cut-off energy.

The antiproton energy spectrum at small atmospheric depth is mainly determined
by the production rate of secondary antiprotons, $\sum_A Q_A(E_A, x, E)$ in Eq.~(\ref{eq:stephX}),
and the tertiary antiprotons are not dominant.
The model calculations based both on box-  and $\delta$-approximations reproduced 
the energy spectrum observed at balloon altitude, 
where secondary antiprotons are dominant over tertiary ones. 
It suggests the production energy spectrum of secondary antiprotons 
was properly treated in our calculation.

At large atmospheric depth, tertiary antiprotons dominate over secondary ones.
The calculation results with the box-approximation show good agreement 
with observed data at mountain altitude and sea level above 1~GeV.
The box-approximation was developed by comparing calculated and observed proton spectra
at sea level and mountain altitude above 1~GeV
in previous works by Bowen and Moats~\cite{bib:BowenMoats}, 
and Stephens~\cite{bib:Stephens1997}. 
The observed spectra were reproduced well by the box-approximation above 1 GeV, 
but below 1 GeV the $\delta$-approximation matches the data more closely. 
Calculation results with the box-approximation
show that the energy spectrum is almost flat in a lower energy region 
both at sea level and mountain altitude. 
However, the energy spectra obtained in the $\delta$-approximation 
decrease below 1~GeV irrespective of observation altitude. 
%The measured spectral shapes are consistent with the $\delta$-approximation below 1 GeV.  
A model calculation 
based on a combination of the box- and $\delta$-approximations 
may therefore reproduce the observed spectra better over a wider energy range
than the simple box- or $\delta$-approximation. 
Some accelerator experimental results indicate 
that the shape of the probability function $\Phi(E,E')$ changes
depending on the initial energy of a projectile particle \cite{bib:Mitsui}.
Measurement of the atmospheric antiproton spectrum with better statistical accuracy
over a wider energy range 
would help to estimate a proper shape of $\Phi(E,E')$.

% -------------------------------------------------------------------
% VII. Conclusion
% -------------------------------------------------------------------
\section{Conclusion}
\label{sec:conclusion}
We measured the atmospheric antiproton spectrum at 4.5 -- 26 g/cm$^2$ and 
at sea level in the kinetic energy range 0.2 -- 3.4 GeV for the first time.

We also referred to our previous observed data 
at Mt. Norikura in 1999 \cite{bib:Sanuki_norikura}
to study the propagation of antiprotons in the atmosphere. 
The energy spectra of antiprotons were calculated 
for the balloon altitude, mountain altitude and sea level.
The model calculations based both on box-  and $\delta$-approximations reproduced 
the energy spectrum observed at balloon altitude, 
where secondary antiprotons are dominant over tertiary ones. 
This suggests 
the production energy spectrum of secondary antiprotons was
properly treated in our calculation.
The spectral shapes of our three measurements below 1~GeV were
reproduced by the $\delta$-approximation,
while the calculated flux amplitude at larger depths and higher energies 
was not well-matched to the data.
The opposite is true for the box-approximation,
which matched the flux data above 1~GeV,
but was not well-matched to the data at larger depths and lower energies.
A model calculation 
based on a combination of the box- and $\delta$-approximations
may reproduce the observed spectra better over a wider energy range
than the simple box- or $\delta$-approximation.

Our measurement of antiproton spectra in the atmosphere
suggests that the shape of the probability function $\Phi(E,E')$ depends 
on the initial energy of the projectile particle.
Measurement of the atmospheric antiproton spectrum with better statistical accuracy
over a wider energy range is highly desirable
to improve accuracy of the model calculations.

\begin{ack}
We would like to thank NASA/GSFC/WFF BPO and NSBF for 
their collaboration and making possible the balloon
expedition. We also thank KEK, ISAS, ICEPP/The University of Tokyo, and 
RESCEU/The University of Tokyo for continuous support. 
This experiment was supported in Japan by KAKENHI
(12047206 and 12047227) from MEXT.
\end{ack}
\clearpage

% -------------------------------------------------------------------
% References
% -------------------------------------------------------------------

% -------------------------------------------------------------------
% Figures
% -------------------------------------------------------------------
\clearpage

\clearpage

\clearpage

\clearpage

% -------------------------------------------------------------------
% Tables
% -------------------------------------------------------------------
\clearpage
\begin{table}
%  \caption{Observed antiproton flux at sea level, 994 g/cm$^2$ at Tsukuba, 1997.}
  \caption{Observed antiproton flux at sea level, 994 g/cm$^2$ at Tsukuba in 1997.}
  \begin{center}
  \vspace{5mm}
  \begin{tabular}{cc|cc|c}
    \hline \hline
    \multicolumn{2}{c|}{Kinetic energy (GeV)} & \multicolumn{2}{c|}{Numbers} & Antiproton flux \\
    range & mean & antiproton & background & (m$^{-2}$ sr$^{-1}$ s$^{-1}$ GeV$^{-1}$) \\ \hline
    0.20$-$0.41 & - & 0 & 0.0 & $8.51\,\times 10^{-5}$ upper limit \\
    0.41$-$0.82 & 0.50 & 2 & 0.0 & $6.02\,^{+6.77}_{-3.79}$$^{+0.10}_{-0.11} \times 10^{-5}$ \\
    0.82$-$1.67 & 1.19 & 13 & 0.1 & $2.10\,^{+0.70}_{-0.60}$$^{+0.07}_{-0.07} \times 10^{-4}$ \\
    1.67$-$3.40 & 2.47 & 10 & 1.9 & $1.11\,^{+0.52}_{-0.44}$$^{+0.12}_{-0.08} \times 10^{-4}$ \\
    \hline
  \end{tabular}
  \end{center}
  \label{tab:flux_gnd}
\end{table}

\begin{table}
%  \caption{Observed antiproton flux at the atmospheric depth of 4.5 - 26 g/cm$^2$ at Ft. Sumner, 2001.}
  \caption{Observed antiproton flux at the atmospheric depth of 4.5 - 26 g/cm$^2$ at Ft. Sumner in 2001.}
  \begin{center}
  \vspace{5mm}
  \begin{tabular}{cc|cc|c}
    \hline \hline
    \multicolumn{2}{c|}{Kinetic energy (GeV)} & \multicolumn{2}{c|}{Numbers} & Antiproton flux \\
    range & mean & antiproton & background & (m$^{-2}$ sr$^{-1}$ s$^{-1}$ GeV$^{-1}$) \\ \hline
    0.20$-$0.35 & 0.20 & 1 & 0.0 & $1.55\,^{+2.71}_{-0.97}$$^{+0.06}_{-0.06} \times 10^{-3}$ \\
    0.35$-$0.62 & 0.54 & 6 & 0.0 & $4.19\,^{+2.29}_{-1.52}$$^{+0.04}_{-0.04} \times 10^{-3}$ \\
    0.62$-$1.09 & 0.89 & 22 & 0.0 & $8.34\,^{+1.78}_{-1.78}$$^{+0.24}_{-0.24} \times 10^{-3}$ \\
    1.09$-$1.93 & 1.53 & 50 & 0.4 & $1.26\,^{+0.18}_{-0.18}$$^{+0.05}_{-0.05} \times 10^{-2}$ \\
    1.93$-$3.40 & 2.61 & 77 & 4.0 & $1.31\,^{+0.15}_{-0.15}$$^{+0.06}_{-0.06} \times 10^{-2}$ \\
    \hline
  \end{tabular}
  \end{center}
  \label{tab:flux_e34}
\end{table}

\end{document}